\documentclass[%
reprint,
superscriptaddress,
amsmath,amssymb,
aps,
pra,
]{revtex4-2}

\makeatletter
\def\NAT@def@citea{\def\@citea{\NAT@separator}}%Remove spaces between multiple citations
\makeatother

\usepackage{setspace}
\usepackage{graphicx}% Include figure files
\usepackage{dcolumn}% Align table columns on decimal point
\usepackage{bm}% bold math
\usepackage[colorlinks=true, allcolors=blue]{hyperref}% add hypertext capabilities
\usepackage[export]{adjustbox}% adjustments to figure alignment
\usepackage{physics}
\usepackage{csquotes}
\usepackage{float}
\usepackage{xcolor}
\usepackage{todonotes}
\usepackage{orcidlink}
\usepackage{cancel}
\usepackage[normalem]{ulem}
\usepackage{enumitem}
\setlist[enumerate]{topsep=1pt,itemsep=-1ex,partopsep=1ex,parsep=1ex}

\usepackage{mathtools}
\usepackage{letltxmacro}
\LetLtxMacro\orgvdots\vdots
\LetLtxMacro\orgddots\ddots

\makeatletter
\DeclareRobustCommand\vdots{%
	\mathpalette\@vdots{}%
}
\newcommand*{\@vdots}[2]{%
	\sbox0{$#1\cdotp\cdotp\cdotp\m@th$}%
	\sbox2{$#1.\m@th$}%
	\vbox{%
		\dimen@=\wd0 %
		\advance\dimen@ -3\ht2 %
		\kern.5\dimen@
		\dimen@=\wd2 %
		\advance\dimen@ -\ht2 %
		\dimen2=\wd0 %
		\advance\dimen2 -\dimen@
		\vbox to \dimen2{%
			\offinterlineskip
			\copy2 \vfill\copy2 \vfill\copy2 %
		}%
	}%
}
\DeclareRobustCommand\ddots{%
	\mathinner{%
		\mathpalette\@ddots{}%
		\mkern\thinmuskip
	}%
}
\newcommand*{\@ddots}[2]{%
	\sbox0{$#1\cdotp\cdotp\cdotp\m@th$}%
	\sbox2{$#1.\m@th$}%
	\vbox{%
		\dimen@=\wd0 %
		\advance\dimen@ -3\ht2 %
		\kern.5\dimen@
		\dimen@=\wd2 %
		\advance\dimen@ -\ht2 %
		\dimen2=\wd0 %
		\advance\dimen2 -\dimen@
		\vbox to \dimen2{%
			\offinterlineskip
			\hbox{$#1\mathpunct{.}\m@th$}%
			\vfill
			\hbox{$#1\mathpunct{\kern\wd2}\mathpunct{.}\m@th$}%
			\vfill
			\hbox{$#1\mathpunct{\kern\wd2}\mathpunct{\kern\wd2}\mathpunct{.}\m@th$}%
		}%
	}%
}
\makeatother

\begin{document}
	
	% \preprint{APS/123-QED}
	
	\title{Native QR Factorization on Programmable Photonic Meshes}
	
	\author{S.A.\,Fldzhyan\,\orcidlink{0000-0002-9174-8019}}
	\email[E-mail me at: ]{fldzhyansa@my.msu.ru}
	\affiliation{%
		Faculty of Physics, M.\,V. Lomonosov Moscow State University, Leninskie Gory 1, Moscow, 119991, Russia
	}%

	\author{S.S.\,Straupe\,\orcidlink{0000-0001-9810-1958}}
	\affiliation{Sber Quantum Technology Center, Kutuzovski prospect 32, Moscow, 121170, Russia}
	\affiliation{%
		Faculty of Physics, M.\,V. Lomonosov Moscow State University, Leninskie Gory 1, Moscow, 119991, Russia
	}%
	
	\author{M.Yu.\,Saygin\,\orcidlink{0000-0001-5494-6801}}
	\affiliation{Sber Quantum Technology Center, Kutuzovski prospect 32, Moscow, 121170, Russia}
	\affiliation{%
		Faculty of Physics, M.\,V. Lomonosov Moscow State University, Leninskie Gory 1, Moscow, 119991, Russia
	}%

	\begin{abstract}
		
		We propose a photonic native procedure for computing the QR factorization of a matrix using a programmable unitary interferometer mesh. The method configures the mesh through a sequence of local power routing steps within tunable two mode interferometric elements, while reading out the resulting upper triangular factor directly from the optical outputs. The number of physical operations grows as $ O(N\log_2N)$ with matrix size $N$, reducing the runtime relative to standard digital QR routines, which scale cubically ($O(N^3)$). Beyond single factorizations, the same architecture supports iterative spectral computations by reusing the configured interferometer in a mirrored arrangement that implements the core update step of the QR eigenvalue algorithm. We also describe related optical procedures for Hessenberg reduction and bidiagonalization, serving as compatible preprocessors for QR and SVD workflows. A comparison with the systolic array computational architecture is provided. Our approach exhibits comparable asymptotic complexity for blocked QR decomposition and is more efficient for Hessenberg reduction and bidiagonalization. %Finally, we analyze the impact of finite control resolution on the computed factors.
		
	\end{abstract}
	
	% \date{\today}
	\def\monthname#1{\ifcase#1\or January\or February\or March\or April\or May\or June\or July\or August\or September\or October\or November\or December\fi}
	\date{\monthname{\the\month}~\the\year}
	
	\maketitle
	% \tableofcontents
	
	\section{Introduction}\label{sec:Intro}
	Programmable integrated photonics has matured into a practical hardware platform for realizing large, reconfigurable linear transformations on optical fields \cite{bogaertsProgrammablePhotonicCircuits2020,wuProgrammableIntegratedPhotonic2024}. This progress spans novel mesh architectures and designs \cite{brugiereNewDesignsLinear2025,marchesinBraidedInterferometerMesh2025,girouardNearoptimalDecompositionUnitary2026}, robust calibration and programming strategies \cite{kondratyevLargescaleErrortolerantProgrammable2024,kondratyevEffectiveProgrammingPhotonic2026,kuzminLeveragingFeaturebasedModel2025,xiaoResidualCalibrationHighprecision2025}, and fundamental studies of algorithmic complexity and physical limits \cite{nemkovComplexityenergyTradeoffProgrammable2025,taguchiSubquadraticScalableApproximate2025,hamerlyInformationtheoreticLimitProgrammable2025,talibPhotonicMatrixMultiplication2025,linRobustMZIBasedOptical2025}. These advances have enabled hardware demonstrations of core linear algebra operations performed directly in the optical domain \cite{tangTwolayerIntegratedPhotonic2022,tangWaveguidemultiplexedPhotonicMatrix2025,milanizadehRecursiveMZIMesh2020,pesericoIntegratedPhotonicTensor2023,chenIterativePhotonicProcessor2022,chenOefficientIterativeMatrix2024,carmonaLuxIALightweightUnitary2025,cavicchioliProgrammableIntegratedPhotonic2024}. Together, they motivate a complementary perspective: rather than using optical meshes solely as analog accelerators for repeated matrix-vector multiplications, we can harness them to physically execute structured linear algebra routines whose conventional digital implementations are bottlenecked by sequential arithmetic.

	A canonical example is the QR factorization. For a matrix $A\in\mathbb{C}^{N\times N}$, QR produces a unitary $Q$ and an upper triangular $R$ such that $A=QR$. QR is central to least squares solvers, orthogonalization, and iterative spectral methods such as the QR algorithm \cite{tyrtyshnikovBriefIntroductionNumerical1997, golubMatrixComputations2013, arbenzLectureNotesSolving2016}. In standard numerical linear algebra, dense QR requires $O(N^3)$ digital floating-point operations, implemented via Householder reflections or Givens rotations. This arithmetic cost, however, is not necessarily aligned with the physical capabilities of a programmable interferometer: once a unitary transformation is configured, optical propagation applies it at essentially constant latency, and the dominant overhead shifts to device configuration and readout. Complex field amplitudes can be measured, for example, using coherent I/Q receivers \cite{khachaturianIQPhotonicReceiver2021}, and intensity-only approaches can reconstruct transfer matrices in many settings \cite{bantyshFastReconstructionProgrammable2023, bantyshFastReconstructionProgrammable2024}. In what follows, we assume
	% parallel coherent readout of all output ports so
	that reading the complex field in one port is an $O(1)$ operation.
	
	In this work we propose an analog QR procedure tailored to universal unitary meshes. The key primitive is a tunable $2\times2$ interferometric block whose phase parameters can be adjusted to \emph{concentrate} optical power from two inputs into a single output, thereby implementing a complex Givens rotation. By composing such local power concentration operations within a triangular Reck mesh \cite{reckExperimentalRealizationAny1994}, the device configures a left unitary transformation $U$ satisfying $UA = R$, where $R$ is upper triangular and can be directly measured, column by column, from the optical outputs. The standard QR factorization then follows as $A=QR$ with $Q:=U^\dagger$. Importantly, the unitary factor is stored in the settings of the mesh. When needed, it can be queried optically using reciprocity and inverting calibrated phases \cite{pottonReciprocityOptics2004}.
	
	Beyond a single factorization, we show that the same hardware can implement the factor swap update of the iterative QR eigenvalue algorithm using mirrored inverse modules (Section~\ref{sec:QR alg}). We also provide compatible optical procedures for Hessenberg reduction and bidiagonalization (Sections~\ref{sec:Hessenberg} and \ref{sec:Bidiagonal}), which are standard preprocessors for efficient QR- and SVD-based spectral computations. In Section~\ref{sec:Systolic} we compare our proposal against the systolic computational architecture \cite{kung1978systolic}. Finally, Section~\ref{sec:Hardware} analyzes finite precision control.

	\section{Optical QR decomposition}\label{sec:QR dec}

	This section presents the core primitive necessary for optical QR decomposition: physically triangularizing a target matrix $A$ by configuring a programmable unitary interferometer. The objective is to realize a unitary left action such that $UA=R$ is upper triangular while directly measuring $R$ during configuration.
	
	\subsection{Physical operation}
	When analyzing the asymptotic complexity of our protocols, we adopt the notion of a \emph{physical operation}, defined as any of the following: (i) encoding an input vector via a physical preparation scheme and injecting it into $N$ ports, (ii) reading out $N$ complex outputs, and (iii) reconfiguring tunable elements of the photonic mesh. Our QR procedure requires $O(N\log_2 N)$ such operations ($N$ column injections and measurements with $O(\log_2 N)$ independent elements reconfigurations each).
	% matching the $O(N^2)$ degrees of freedom of a dense matrix and not exceeding the loading cost for X-bar processors \cite{chilesDesignFabricationMetrology2018, xuParallelOpticalCoherent2022, giamougiannisNeuromorphicSiliconPhotonics2023}.

	\subsection{Tunable $2\times2$ block and power concentration}
\begin{figure}[hbt!]
		\centering
		\includegraphics[width=\linewidth]{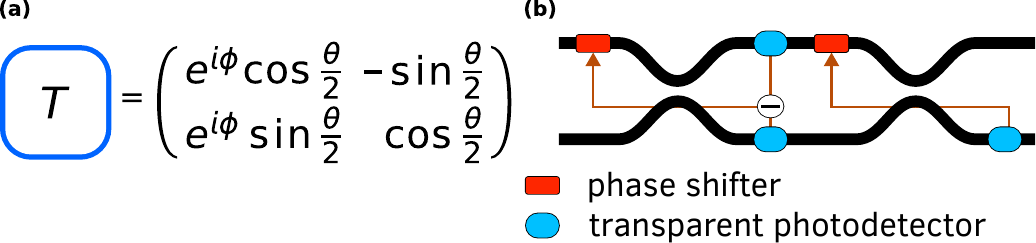}
		\caption{ \textbf{Self-configuring optoelectronic concentrator block}: (a) Fundamental tunable $2\times2$ interferometric block $T$. (b) One possible implementation of the tunable $2\times2$ block using a MZI with local feedback. The phase shifters are configured to null the detected signal, thereby zeroing the second output port.}
		\label{fig:2.05}
	\end{figure}

	The procedure relies on a tunable $2\times2$ unitary block $T$, presented in Fig.~\ref{fig:2.05}(a), that independently controls a splitting ratio and a relative phase. This block performs what is called a Givens rotation. Such blocks can physically be realized, e.g., using a phase shifter combined with a tunable beam splitter, a standard Mach-Zehnder interferometer (MZI), 3-MZI, or an MZI with a crossing \cite{reckExperimentalRealizationAny1994, clementsOptimalDesignUniversal2016, hamerlyAsymptoticallyFaulttolerantProgrammable2022}. The sole requirement is power concentration capability: given an arbitrary input vector $(x_1,x_2)^T$, one can e.g. tune the phase $\phi=\arg(-x_2/x_1)$ and splitting ratio $\tan(\theta/2)=|x_2/x_1|$ to obtain
	\begin{equation}\label{eq:concentrate 2}
		\begin{pmatrix}
			e^{i\phi}\cos{\frac{\theta}{2}} & -\sin{\frac{\theta}{2}}\\
			e^{i\phi}\sin{\frac{\theta}{2}} & \cos{\frac{\theta}{2}}
		\end{pmatrix}\!\!
		\begin{pmatrix}
			x_1\\
			x_2
		\end{pmatrix}=
		\begin{pmatrix}
			-e^{i\arg{x_2}}\sqrt{|x_1|^2+|x_2|^2}\\
			0
		\end{pmatrix}.
	\end{equation}
	Equation~\eqref{eq:concentrate 2} is the optical analogue of a complex Givens rotation: one component is nulled while preserving total power (up to a phase).
	
	The physical setup of a unitary $T$ block we envision is illustrated in Fig.~\ref{fig:2.05}(b) (see, e.g., \cite{ditriaHighPrecisionAutomatedSetting2025}). In this implementation, nulling the second output channel is achieved solely through local feedback. Under this assumption, we assign a complexity of $O(1)$ to such a concentration operation within a single block.

	\subsection{Reck mesh and concentrators}
	
\begin{figure}[hbt!]
		\centering
		\includegraphics[width=\linewidth]{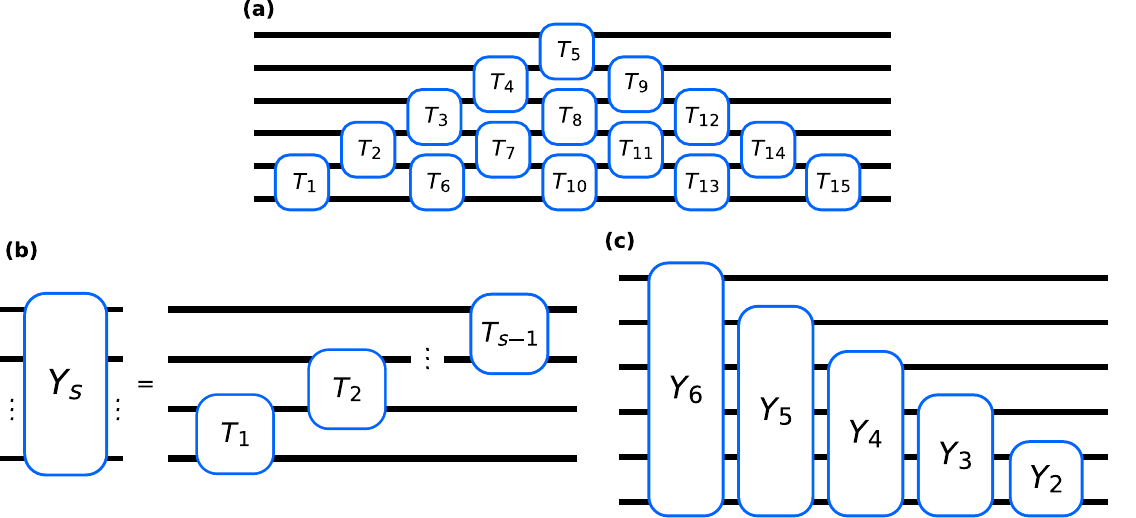}
		\caption{\textbf{Relevant optical interferometers}: (a) Reck universal unitary mesh~\cite{reckExperimentalRealizationAny1994} for $N=6$. (b) Scheme of Givens  ladder concentrator $Y_s$. (c) Composite concentrating scheme.}
		\label{fig:2.1}
	\end{figure}

	We adopt the triangular Reck decomposition mesh as the universal unitary backbone (Fig.~\ref{fig:2.1}(a)) \cite{reckExperimentalRealizationAny1994}. Following \cite{russellDirectDiallingHaar2017}, we group interferometric blocks into \emph{concentrators} $Y_s$, e.g. formed from Givens rotation ladder as shown in Fig.~\ref{fig:2.1}(b). This yields the compact representation in Fig.~\ref{fig:2.1}(c). By sequentially applying the power concentration property within the constituent $T$ blocks, the total optical power of any input incident on $Y_s$ can always be routed into the uppermost channel. A notable feature of this scheme is its capacity for self-configuration \cite{millerSelfconfiguringUniversalLinear2013}: the system can align itself using only local feedback loops that maximize power at designated nodes, potentially eliminating the need for external digital processing for unitary matrix encoding.
	
\begin{figure}[hbt!]
		\centering
		\includegraphics[width=0.6\linewidth]{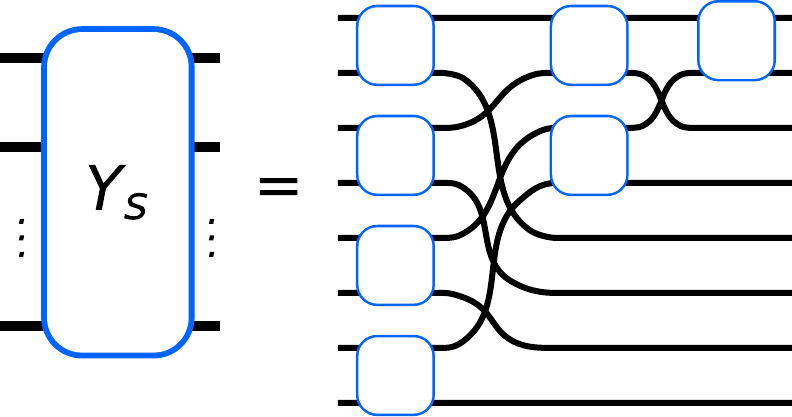}
		\caption{\textbf{Example of the logarithmic tree concentrator photonic circuit $\boldsymbol{Y_s}$ for $\boldsymbol{s=8}$}. The structure routes any input vector to the top output by configuring each $T$ block.}
		\label{fig:2.15}
	\end{figure}

	In this work we adopt a more compact implementation of the concentrator $Y_s$. The device shown in Fig.~\ref{fig:2.15} (shown for $s=8$ as an example) is based on a logarithmic tree and can also concentrate power into the first channel. Its advantage over the Givens ladder in Fig.~\ref{fig:2.1}(b) is that more $T$ blocks can be configured in parallel rather than sequentially, so configuring $Y_s$ takes $O(\lceil\log_2 s\rceil)$ steps. In the following discussion we omit the ceiling notation for simplicity.

	\subsection{Optical QR decomposition protocol}
	
\begin{figure}[hbt!]
		\centering
		\includegraphics[width=\linewidth]{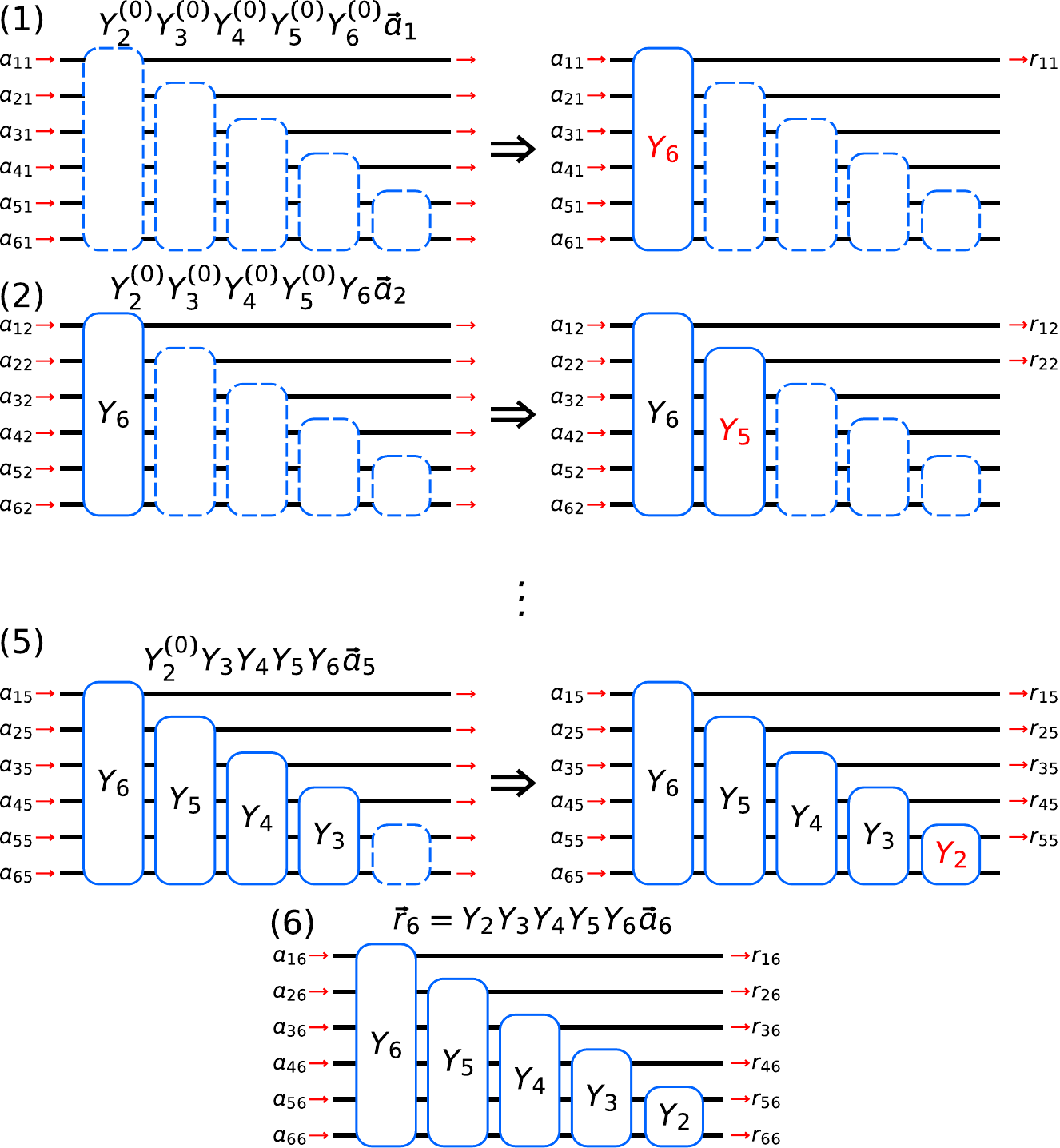}
		\caption{\textbf{Illustration of the physical QR decomposition for $N=6$}. Columns of $A$ are injected sequentially while concentrators $Y_N$ through $Y_2$ zero subdiagonal elements. Red symbols mark the newly configured concentrator $Y_{N+1-i}$ at step $(i)$. The final injection yields the last column of $R$.}
		\label{fig:2.2}
	\end{figure}

	Let $A=[\vec a_1,\ldots,\vec a_N]$ be the target matrix provided as a sequence of input columns. The optical QR procedure configures the mesh so that after processing $\vec a_i$ the output has zeros below row $i$, and the resulting output column becomes the $i$-th column of $R$. Fig.~\ref{fig:2.2} illustrates the step by step procedure for $N=6$. The \emph{optical QR decomposition} proceeds as follows:
	
	\begin{enumerate}
		\item \textbf{Initialization:} Initialize all concentrators, i.e., set all $Y^{(0)}_s$  (depicted as white boxes with dashed borders in Fig.~\ref{fig:2.2}). Set the column index to $i=1$.
		\item \textbf{Injection:} Inject the column vector $\vec a_i$.
		\item \textbf{Configure the concentrator:} Configure the concentrator $Y_{N+1-i}$ to concentrate optical power into the first $i$ output ports and measure the output. This requires $O(\log_2{(N-i)})$ steps.
		\item \textbf{Iterate:} Increment $i$ and repeat steps 2 \& 3 until $i=N$.
		\item \textbf{Final column:} Inject $\vec a_N$ to read out the last column of $R$.
	\end{enumerate}
	
	At the conclusion of the process, the programmed unitary $U$ satisfies
	\begin{equation}
		UA = R,
	\end{equation}
	where $R$ is upper triangular and is obtained from the measured output columns. For the $N=6$ example in Fig.~\ref{fig:2.2}, the resulting $R$ has the form
	\begin{equation}
		R=\begin{pmatrix}
			r_{11} & r_{12} & r_{13} & r_{14} & r_{15} & r_{16}\\
			0 & r_{22} & r_{23} & r_{24} & r_{25} & r_{26}\\
			0 & 0 & r_{33} & r_{34} & r_{35} & r_{36}\\
			0 & 0 & 0 & r_{44} & r_{45} & r_{46} \\
			0 & 0 & 0 & 0 & r_{55} & r_{56}\\
			0 & 0 & 0 & 0 & 0 & r_{66}\\
		\end{pmatrix}.
	\end{equation}
	The number of physical operations is $O(N\log_2 N)$: there are $N$ column injections and $\sim O(\log_2 N)$ local configuration steps per column.
	
	\subsection{Recovering $Q$}
	
	Defining $Q:=U^\dagger$ yields the standard QR factorization
	\begin{equation}
		A = QR.
	\end{equation}
	The entries of $Q$ are not explicitly computed during the procedure, instead, $Q$ is stored in the mesh settings. If an application requires explicit access to $Q$ (or the ability to apply $Q$ to vectors), there are two natural options. First, one may characterize the configured unitary $U$ by injecting basis vectors and measuring outputs, which costs $O(N)$. Second, one may exploit optical reciprocity \cite{pottonReciprocityOptics2004}: by reversing the propagation direction and inverting the calibrated phases in the $2\times2$ blocks, the hardware can implement transposition and conjugation relations needed to realize $U^\dagger$ optically. In many routines, including QR-based eigenvalue iterations, it suffices to apply $Q$ and $Q^\dagger$ physically rather than to list their entries.
	
	\subsection{Input preparation}% and multiplexing}

\begin{figure}[hbt!]
	\centering
	\includegraphics[width=0.7\linewidth]{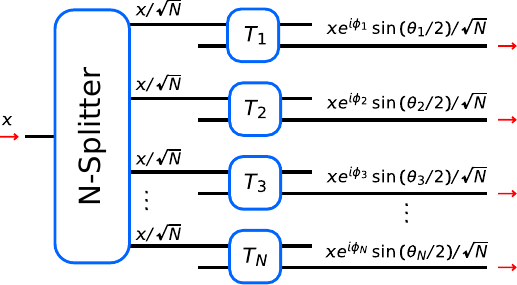}
	\caption{\textbf{Input column vector encoding via tunable MZI blocks}.}
	\label{fig:2.3}
\end{figure}

The column injection stage does not require a full universal encoding of $A$ (X-bar \cite{chilesDesignFabricationMetrology2018, xuParallelOpticalCoherent2022, giamougiannisNeuromorphicSiliconPhotonics2023}, SVD \cite{millerSelfconfiguringUniversalLinear2013} or embedding \cite{tangLowerdepthProgrammableLinear2024, fldzhyanLowdepthTwounitaryDesign2026}) configured for all $N^2$ entries. Instead, a dedicated vector synthesis circuit can generate an arbitrary complex column $\vec a_i$. E.g., scheme in Fig.~\ref{fig:2.3} can prepare any input vector. By launching light into  equal N-splitter and choosing  $T$ settings to match desired amplitudes and phases. As we have outlined before, we assume that the vector preparation is done in $O(1)$.

\section{Optical implementation of the QR eigenvalue iteration}\label{sec:QR alg}

Having an $O(N\log_2 N)$ optical method to obtain $UA=R$, one can reuse it repeatedly to compute spectral information rather than a single factorization. The QR algorithm generates a sequence $\{A_k\}$ by factoring each iterate and swapping the factors,
\begin{equation}
	A_k = Q_k R_k, \qquad A_{k+1} = R_k Q_k = Q_k^\dagger A_k Q_k,
\end{equation}
so the iteration preserves eigenvalues while driving the matrix toward Schur form (upper triangular or diagonal for normal matrices) by unitary similarity transformations \cite{tyrtyshnikovBriefIntroductionNumerical1997, golubMatrixComputations2013, arbenzLectureNotesSolving2016}. With shifts, convergence is accelerated in practice, and for normal matrices, in the limit, eigenvectors are obtained from the accumulated $Q_k$ factors. Here we explain how the same photonic hardware can implement
that swap natively via mirrored modules, turning each iteration into a reprogram and measure loop.

\subsection{Mirrored modules and factor swapping in hardware}

\begin{figure}[hbt!]
	\centering
	\includegraphics[width=1\linewidth]{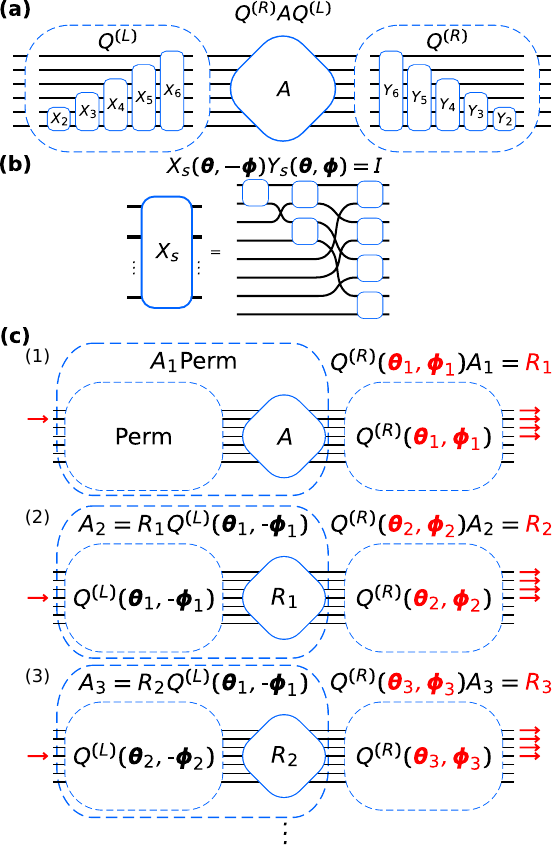}
	\caption{\textbf{Illustration of the optical iterative QR algorithm for $\boldsymbol{N=6}$}. (a) Two universal interferometers with mirrored inverse concentrators ($Q^{(L)}$, $Q^{(R)}$). (b) Concentrator $X_s$, the structural inverse of $Y_s$. (c) First three iterations, red symbols mark parameters determined at step $(i)$.}
	\label{fig:3}
\end{figure}

The optical QR procedure yields a left factorization $U_k A_k = R_k$ with unitary $U_k$, so $Q_k = U_k^\dagger$. To implement the update $A_{k+1} = R_k Q_k$ optically, we use two mirrored Reck-based programmable units, $Q^{(R)}$ and $Q^{(L)}$, flanking a central X-bar array (Fig.~\ref{fig:3}(a)). The X‑bar is essential for maintaining $O(N\log_2 N)$ iteration complexity, as it can be configured in $O(N)$ or even $O(1)$ time depending on the architecture \cite{nemkovComplexityenergyTradeoffProgrammable2025}. When programming mirrored units with opposite phase and same splitting settings, they realize inverse transformations $Q^{(L)}(\bm{\theta}, -\bm{\phi})\,Q^{(R)}(\bm{\theta}, \bm{\phi}) = I$. Structurally, $Q^{(L)}$ employs concentrators $X_s$ that reverse the operation order of the $Y_s$ concentrators in $Q^{(R)}$ (Fig.~\ref{fig:3}(b)), enabling a native factor swap: after measuring $R_k$, it is reencoded in the central stage and right-multiplied by $U_k^\dagger$ via the mirrored module to produce the next iterate.

\subsection{QR eigenvalue iteration protocol}

The iterative \emph{optical QR algorithm} proceeds as follows:

\begin{enumerate}
	\item \textbf{Initialization:} Initialize $Q^{(L)}$ and $Q^{(R)}$, load $A_1 = A$ into the central X-bar array, and set $i=1$.
	\item \textbf{QR step:} Inject columns $j=1,\dots,N$ of $A_i$ to execute the optical QR decomposition (Sec.~\ref{sec:QR dec}), yielding parameters $\bm{\theta_i}$ and $\bm{\phi_i}$ that configure $Q^{(R)}(\bm{\theta}_i,\bm{\phi}_i)$ and the measured upper triangular $R_i$, such that $Q^{(R)}(\bm{\theta}_i,\bm{\phi}_i) A_i = R_i$.
	\item \textbf{Factor swap:} Form $A_{i+1} = R_i Q^{(R)\dagger}(\bm{\theta}_i,\bm{\phi}_i)$ by loading $R_i$ into the X-bar array and programming $Q^{(L)}$ with inverted phases $-\bm{\phi}_i$. Because $Q^{(L)}(\bm{\theta}_i,-\bm{\phi}_i) = Q^{(R)\dagger}(\bm{\theta}_i,\bm{\phi}_i)$, this implements the factor swap in hardware.
	\item \textbf{Iterate:} Increment $i$ and repeat steps 2 \& 3 until $A_i$ converges to Schur form.
\end{enumerate}
The data flow for the first three iterations is conceptually mapped out in Fig.~\ref{fig:3}(c). Setting $\bm{\phi}=\bm0$ and $\bm{\theta}=\bm0$ during initialization of $Q^{(L)}$ encodes a permutation, which only needs to be accounted for at the first step ($i=1$) by appropriately choosing the input channels.

Since the QR factorization is performed column by column, some hardware simplifications are possible. For example, replacing the full left module with a single column preparation in Fig.~\ref{fig:2.3} can reduce optical depth. In that case, one must ensure that at each internal step the incident distribution matches the required column of $Q^{(R)\dagger}$. If $Q^{(R)}$ has been characterized, the needed physical parameters can be set in $O(1)$ time per column.

The proposed scheme can be readily modified to implement the shifted QR algorithm, a standard technique for accelerating convergence \cite{arbenzLectureNotesSolving2016}. This extension is straightforward because applying a shift manifests as a simple correction to a single component of the output vector for each injected column input. For a target matrix $A$ preprocessed to Hessenberg form (see next section), the implicit Q theorem guarantees that a single pair of concentrators (in the Givens ladder form) $Q^{(R)}=X_N$, $Q^{(L)}=Y_N$ is sufficient to perform QR iterations with shifts using the ``chasing the bulge'' technique \cite{golubMatrixComputations2013} (Theorem 7.4.2 and Paragraph 8.3.5). This greatly simplifies the optical scheme.

\section{Optical Hessenberg reduction}\label{sec:Hessenberg}

In standard dense matrix eigensolvers, QR iterations are rarely applied to $A$ directly. Instead, $A$ is reduced to upper Hessenberg form via a unitary similarity transform, a preprocessing step that costs $O(N^3)$ arithmetic operations in digital implementations \cite{golubMatrixComputations2013}. One constructs a unitary $U$ such that
\begin{equation}
	U A U^\dagger =H
	% =\left(\begin{smallmatrix}
		%         h^{1}_{1} & h^{1}_{2} & h^{1}_{3}&\cdots & &  \cdots& h^{1}_{N}\\
		%         h^{2}_{1} & h^{2}_{2} & h^{2}_{3}&\ddots & & & \vdots\\
		%         0 & h^{3}_{2} & h^{3}_{3} &\ddots & &  & \\
		%          & 0 & h^{4}_{3} &\ddots & &  & \\
		%          &  & & \ddots& \ddots & \ddots & \vdots\\
		%           \vdots&   & & 0& h^{N-1}_{N-2}& h^{N-1}_{N-1} & h^{N-1}_{N}\\
		%         0 & \cdots & & & 0& h^{N}_{N-1} & h^{N}_{N}
		%     \end{smallmatrix}\right)
	,
\end{equation}
where $H$ is upper Hessenberg: all entries below the first subdiagonal are zero, i.e., $h_{ij}=0$ for $i>j+1$. The key benefit is that QR iterations on a Hessenberg matrix can be performed in $O(N^2)$ digital steps using bulge chasing, rather than $O(N^3)$ \cite{tyrtyshnikovBriefIntroductionNumerical1997, golubMatrixComputations2013}.

\begin{figure}[hbt!]
	\centering
	\includegraphics[width=\linewidth]{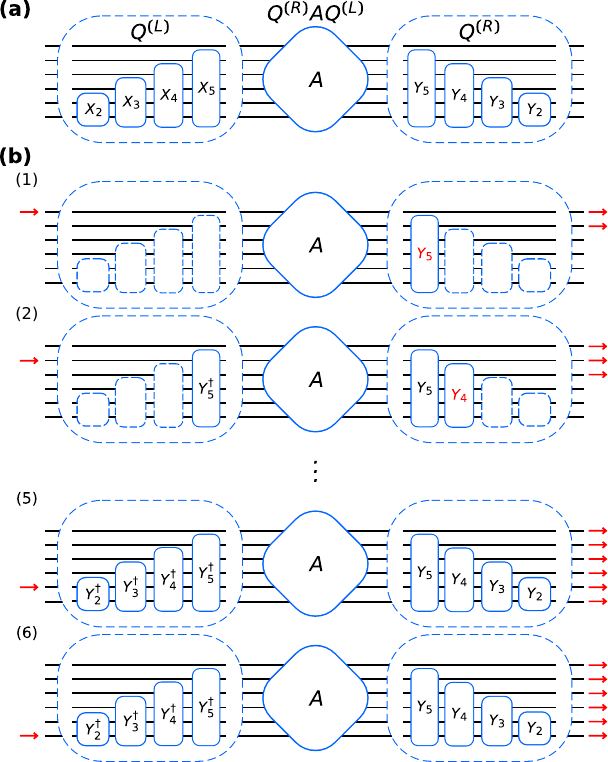}
	\caption{\textbf{Illustration of the optical Hessenberg decomposition for $\boldsymbol{N=6}$}. (a) Two shortened (nonuniversal) interferometers with mirrored inverse concentrators ($Q^{(L)}$, $Q^{(R)}$). (b) Decomposition workflow: red symbols mark the concentrators updated at step $(i)$.}
	\label{fig:4}
\end{figure}

The optical setup in Fig.~\ref{fig:4} implements Hessenberg reduction in $O(N\log_2 N)$ physical operations. Compared to the full QR eigenvalue iteration setup, the unitary modules can be shortened because full universality is not required: only the concentrators involved in eliminating entries below the first subdiagonal need to be configured. The \emph{optical Hessenberg reduction protocol} proceeds as follows:

\begin{enumerate}
	\item \textbf{Initialization:} {Initialize} shortened unitary modules $Q^{(L)}$, $Q^{(R)}$, load $A$ into the central unit, and set $i=1$.
	\item \textbf{Concentration:} Inject light into port $i$, configure $Y_{N-i}$ in $Q^{(R)}$ to concentrate power into the first $i+1$ outputs. The measured output gives column $i$ of $H$.
	\item \textbf{Update:} Program $X_{N-i} = Y_{N-i}^\dagger$ in $Q^{(L)}$ using the same splittings and inverted phases.
	\item \textbf{Iteration:} Increment $i$, repeat steps 2 \& 3 until $i = N-1$.
	\item \textbf{Final colums:} Inject ports $N-1$ and $N$ to obtain the last two columns of $H$.
\end{enumerate}

This optical Hessenberg reducer can serve as a preprocessor: it produces a structured matrix suitable for subsequent optical or numerical QR iterations, with the advantage that the preprocessing avoids $O(N^3)$ cost.

\section{Optical Bidiagonalization}\label{sec:Bidiagonal}
While Hessenberg reduction prepares a matrix for QR algorithm, many applications instead require singular values and singular vectors, where the standard entry point is bidiagonalization \cite{tyrtyshnikovBriefIntroductionNumerical1997, golubMatrixComputations2013}. For any matrix $A$, one can find unitary matrices $U$ and $V$ in $O(N^3)$ digital steps such that
\begin{equation}
	UAV=B
	=
	\begin{pmatrix}
		d_1 & f_1 & 0 & & \cdots & 0\\
		0 & d_2 & f_2 & & & \vdots\\
		& & \ddots & \ddots &  & \\
		& & & d_{N-2}&f_{N-2} & 0\\
		\vdots &  & & 0& d_{N-1} & f_{N-1}\\
		0 & \cdots & & & 0 & d_{N}
	\end{pmatrix}
	,
\end{equation}
where $B$ is called (upper) bidiagonal, and $d_i$, $f_i$ can always be chosen real. Subsequent SVD computations are then simplified because $B^TB$ is symmetric tridiagonal.

\begin{figure}
	\centering
	\includegraphics[width=\linewidth]{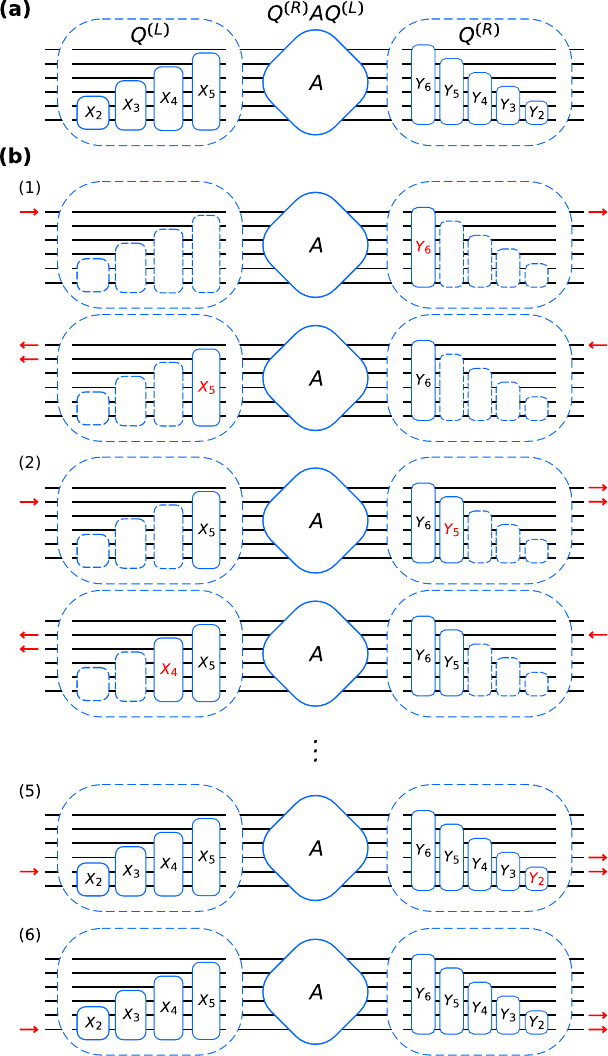}
	\caption{\textbf{Illustration of the optical bidiagonalization for $\boldsymbol{N=6}$}. (a) Setup with one shortened (nonuniversal) and one universal multiport interferometers: $Q^{(L)}$ and $Q^{(R)}$ use mirrored inverse concentrators. (b) Decomposition workflow: red symbols mark the concentrators updated at step $(i)$.}
	\label{fig:5}
\end{figure}

The optical setup in Fig.~\ref{fig:5} performs bidiagonalization in $O(N\log_2 N)$ physical operations by alternating left to right and right to left concentration steps. Optical reciprocity is used by injecting light from the opposite side to reverse the effective order of operations \cite{pottonReciprocityOptics2004}. The \emph{optical bidiagonalization protocol} proceeds as follows:

\begin{enumerate}
	\item \textbf{Initialization:} Initialize $Q^{(L)}$ and $Q^{(R)}$, load $A$ into the central unit, and set $i=1$.
	\item \textbf{Left to right concentration:} Inject port $i$ on the left, configure $Y_{N+1-i}$ in $Q^{(R)}$ to concentrate power into the first $i$ right outputs. The measured output gives column $i$ of $B$.
	\item \textbf{Right to left concentration:} Inject port $i$ on the right, configure $X_{N-i}$ in $Q^{(L)}$ to concentrate power into the first $i+1$ left outputs. The measured output gives row $i$ of $B$.
	\item \textbf{Iteration:} Increment $i$, repeat steps 2 \& 3 until $i = N-1$, then perform one final left to right concentration at $i = N-1$.
	\item \textbf{Final entries:} Inject left port $N$ to obtain the last column of $B$.
\end{enumerate}

The proposed optical bidiagonalization procedure offers an advantage over the optical QR eigenvalue iteration and Hessenberg reduction in terms of locality. In the previously described procedures, physical parameters from the $Y_i$ concentrators had to be transferred to the corresponding $X_i$ concentrators. For Hessenberg reduction, configuration and transfer can occur concurrently, yet the transfer itself remains nonlocal. As shown here, bidiagonalization does not require such parameter transfer and can therefore be configured entirely locally.

\section{Comparison with systolic arrays}\label{sec:Systolic}

To evaluate the proposed optical scheme against a strong hardware baseline, we compare it with systolic arrays~\cite{kung1978systolic, kungVLSIArrayProcessors1985, kurzakDesignImplementationPULSAR2017}. They are widely used today to accelerate various linear algebraic workloads \cite{jouppiInDatacenterPerformanceAnalysis2017,lu2025systolicupdateschemeovercome}.  This comparison is particularly relevant because systolic architectures are known to provide a highly efficient implementation of QR decomposition \cite{gentlemanMatrixTriangularizationSystolic1982, bojanczykNumericallyStableSolution1984}: by mapping Givens rotation updates onto a regular pipelined array, they avoid much of the sequential overhead that limits conventional digital algorithms. Thus, systolic arrays efficiently exploit the structure of QR decomposition to achieve better performance, making them an appropriate reference point for identifying the advantages and limitations of the photonic approach.

\begin{figure}
	\centering
	\includegraphics[width=\linewidth]{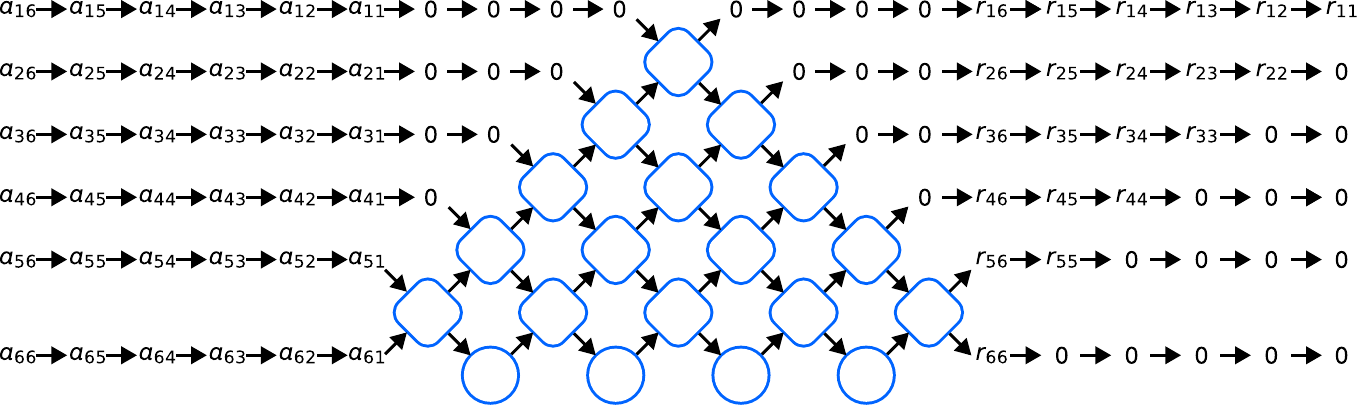}
	\caption{\textbf{Pipelined implementation of QR decomposition on a systolic array proposed in~\cite{bojanczykNumericallyStableSolution1984}}. Each rhombus is configured by the first nonzero input it receives and then stays constant; circles act as passive routing elements.}
	\label{fig:6.1}
\end{figure}

A systolic implementation of QR decomposition using Givens rotations, as presented in \cite{bojanczykNumericallyStableSolution1984}, has a structure illustrated in Fig.~\ref{fig:6.1} that is similar to the Reck mesh in Fig.~\ref{fig:2.1}(a). In this architecture, rhombus cells apply a Givens rotation to two inputs (as in Fig.~\ref{fig:2.05}(a)), while circles simply reroute signals. When a nonzero input signal first reaches a rhombus, the cell configures itself to nullify the output signal in the lower channel and then keeps those settings unchanged. A single column propagates through the array in $2N+1$ steps (the base of the triangle), so the last column emerges after $3N+1$ steps. Thus, the entire factorization completes in $O(N)$ steps which is asymptotically faster than introduced optical $O(N\log_2N)$ protocols. This is because a pipelined architecture can perform configuration and data propagation concurrently, whereas in our optical system these occur sequentially.

We first observe that the optical setup can in principle emulate a pipelined systolic implementation by introducing appropriate delays. One might also try and pursue a concentrator block that achieves power concentration in $O(1)$ steps to approach the same asymptotic. That said, the optical approach exhibits the same asymptotic complexity for blocked implementations in which the device dimension $D$ is much smaller than the full matrix size $N$ ($D \ll N$). Where there is a need for a lot of matrix-vector multiplications with the same matrix.

\begin{figure}
	\centering
	\includegraphics[width=0.9\linewidth]{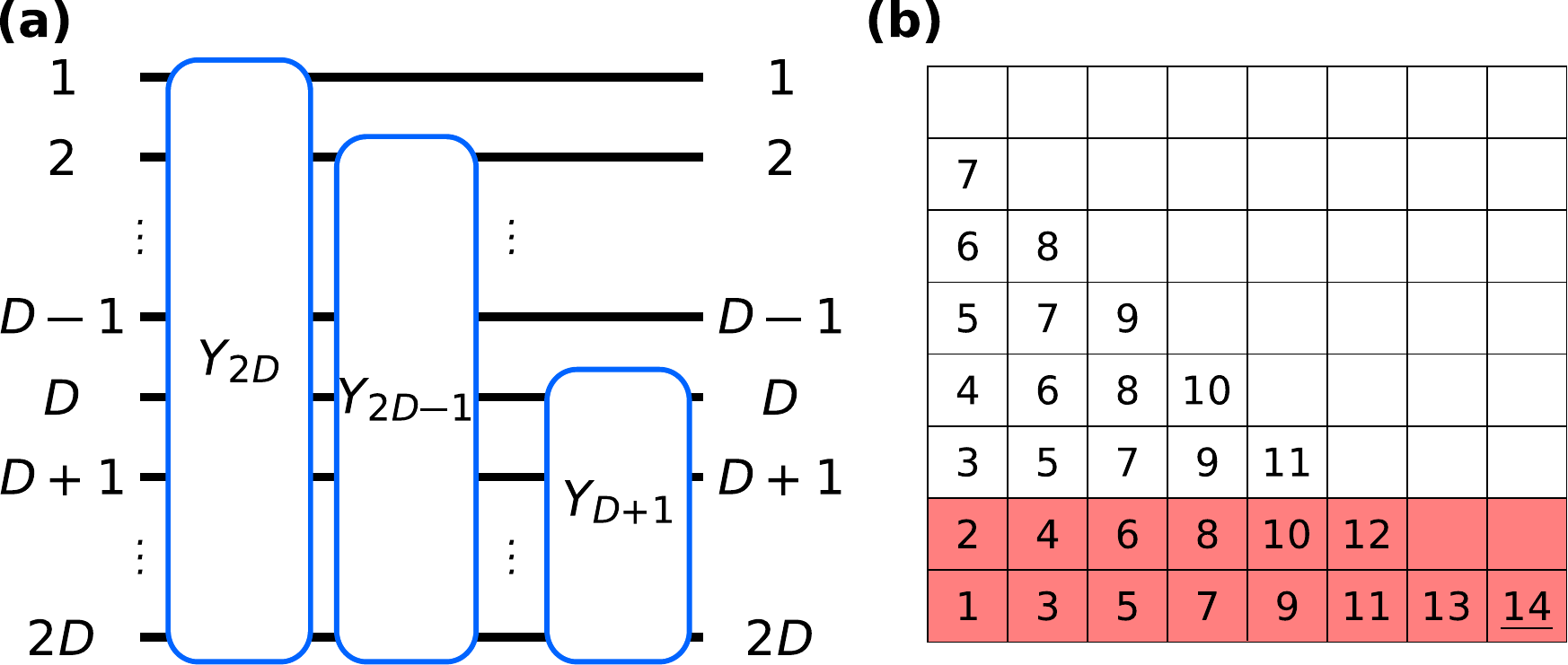}
	\caption{\textbf{Block QR decomposition}. (a) The unitary $Z_D$ performs QR decomposition of any $2D \times D$ matrix. (b) Block QR decomposition order for $N/D = 8$; numbers indicate the nulling order, and blocks with the same number can be nulled in any order. The highlighted $2D \times N$ strip corresponds to the first nulled block. QR decomposition of the last bottom right block is performed with padding.}
	\label{fig:6.2}
\end{figure}

To analyze the asymptotic complexity of this blocked QR decomposition, we use Gaussian pairwise elimination with block iteration \cite{tiskinCommunicationefficientParallelGeneric2007}. To that end, we introduce a unitary matrix $Z_D$ of size $2D \times 2D$ that can perform QR decomposition on any rectangular $2D \times D$ matrix as follows:
\begin{equation}
	Z_D
	\begin{pmatrix}
		A_{1}\\
		A_2
	\end{pmatrix}=
	\begin{pmatrix}
		R_{1}\\
		0
	\end{pmatrix}.
\end{equation}
A physical implementation of $Z_D$ is shown in Fig.~\ref{fig:6.2}(a).

Partition $A$ into $D \times D$ blocks. The $Z_D$ unit nulls a block by operating on a $2D \times N$ strip \cite{tiskinCommunicationefficientParallelGeneric2007} (Fig.~\ref{fig:6.2}(b)). This takes $O(D \log_2 D)$ configuration steps (or $O(D)$ for a systolic array) followed by $O(N)$ vector passes through the configured transform. The per block cost $O(D \log_2 D + N)$ is dominated by the $O(N)$ term for both realizations. Nulling $O(N^2/D^2)$ blocks yields $O(N^3/D^2)$ total complexity. Moreover, because some blocks can be nulled independently, the algorithm can be parallelized across multiple physical units.

For the optical implementation, the dominant step of passing many vectors through the same transformation can be accelerated by higher frequency operation \cite{li2025allopticalcomputing100ghzclock} or by multiplexing $M$ modes (e.g., wavelengths \cite{Feldmann_2021}), reducing the effective cost of the $O(N)$ vector passes by a factor of $M$. This gives a practical speedup that lowers the constant factor and yields an asymptotic complexity of $ O(N^3 / M D^2)$.

Regarding the proposed Hessenberg reduction and bidiagonalization procedures, a naive systolic implementation (analogous to Fig.~\ref{fig:6.1}) incurs $O(N^2)$ complexity since each Givens ladder $Y_s$ requires $O(s)$ operations for both configuration and data propagation, whereas the optical approach achieves $O(N \log_2 N)$. Even implementing the logarithmic tree concentrator from Fig.~\ref{fig:2.15} in a systolic architecture does not recover the asymptotic complexity of the optical approach. The key difference is that data propagation through a systolic array is not instantaneous. Consequently, for the Hessenberg reduction and bidiagonalization described in Sections~\ref{sec:Hessenberg} and \ref{sec:Bidiagonal}, a delay is required before a new input can be injected. This overhead ultimately degrades the asymptotic complexity. In contrast, optical signal propagation is assumed to be instantaneous.

\section{Hardware Imperfections}\label{sec:Hardware}
The algorithms above assume an ideal programmable unitary acting on complex amplitudes, but real devices face loss, finite control precision, and physical layout limits. Here we translate algorithmic steps into hardware requirements by analyzing the effect of phase imprecision.

%\subsection{Bit Precision}
The finite precision of the phase settings arises from inherent noise in the analog concentration scheme shown in Fig.~\ref{fig:2.05}(b). To model this noise, we quantize the phase $\phi$ and the splitting parameter $\theta$ in steps of $2\pi/2^b$ and $\pi/2^b$, respectively, where $b$ is the number of bits representing the noise level. We then choose the quantized parameters to best recover the input vectors and achieve maximal concentration. Note that simply rounding the ideal parameters from \eqref{eq:concentrate 2} is not always optimal.

Column preparation is also subject to finite precision. Fig.~\ref{fig:2.3} illustrates column preparation strategy: input $x$ is chosen so the largest amplitude matrix coefficient is encoded exactly (at $\theta = \pi$), with $\phi_i$ and $\theta_i$ as the tunable parameters.

\begin{figure}[hbt!]
	\centering
	\includegraphics[width=\linewidth]{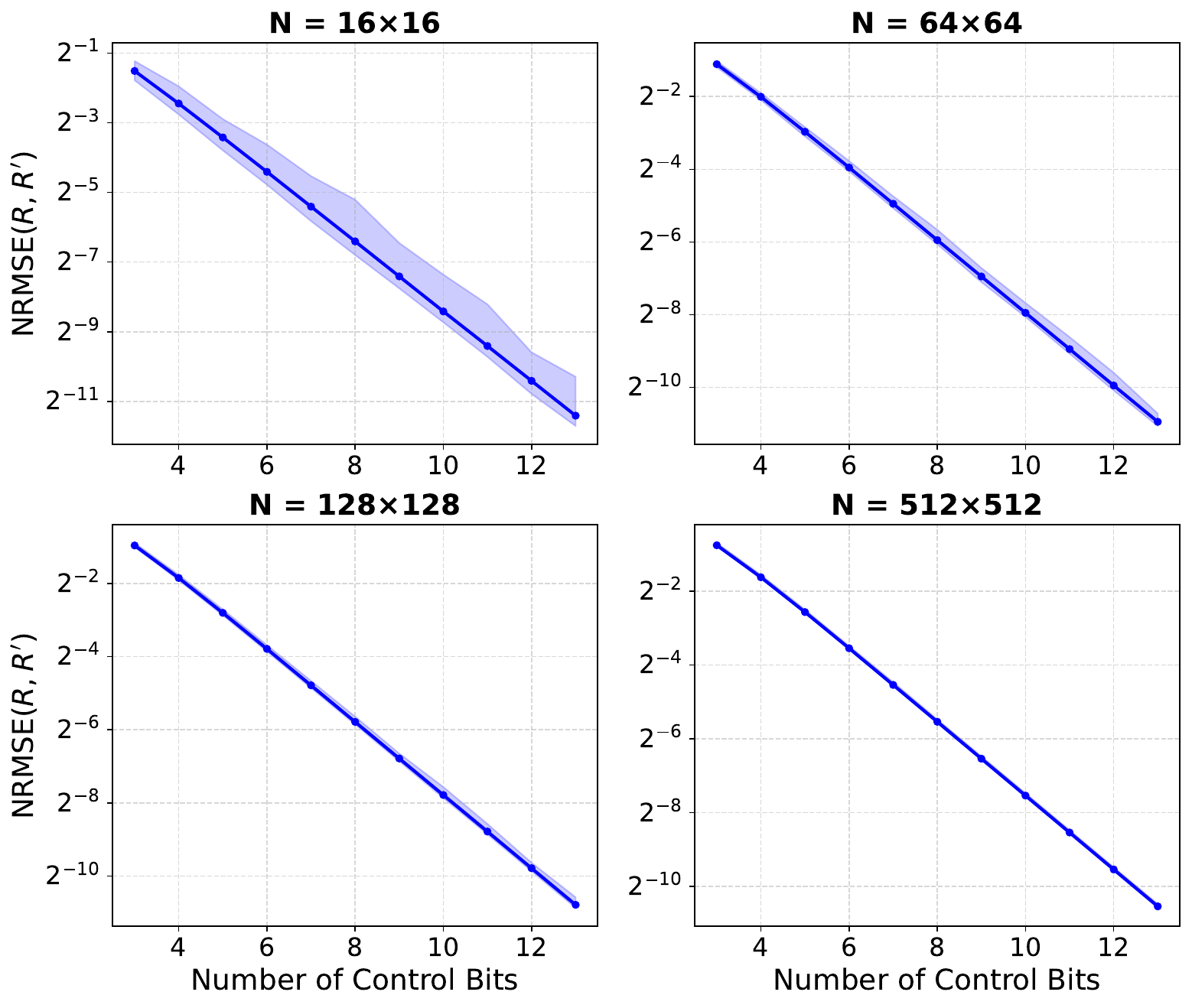}
	\caption{\textbf{Finite precision error analysis of optical QR decomposition}. Normalized root mean square error (NRMSE) of the optical QR result $R'$ versus the ideal $R$, across matrix sizes $N$ and bits of physical precision. Shaded regions: 500 random Ginibre matrices, normalized by the largest coefficient magnitude; solid lines: mean values.}
	\label{fig:7.1}
\end{figure}

Decomposition quality is evaluated by comparing the optical result ($Q', R'$) to a high precision numerical QR ($Q, R$) from SciPy \cite{2020SciPy-NMeth}, using target matrices taken from the Ginibre ensemble normalized by the largest element modulus. To account for nonuniqueness of QR decomposition, we apply a diagonal unitary $\operatorname{diag}(e^{i\arg(\mathrm{diag}[R'R^\dagger])})$ to align phases of $R'$ to $R$ before computing the normalized root mean square error:
\begin{equation}
	\mathrm{NRMSE}(R, R') = 
	\frac{\sqrt{\frac{1}{N^2}\sum_{i,j} |r^{'}_{ij} - r_{ij}|^2}}
	{\frac{2}{N(N+1)}\sum_{i \leq j} |r_{ij}|}.
\end{equation}
Results in Fig.~\ref{fig:7.1} show that high fidelity decomposition is feasible with current experimental capabilities \cite{ditriaHighPrecisionAutomatedSetting2025}.

\section{Discussion}

We have introduced a native analog procedure for QR factorization on programmable unitary photonic meshes. The central idea is to use local power concentration operations in tunable $2\times2$ interferometric blocks to configure a unitary transformation $U$ such that $UA=R$ is upper triangular, while the entries of $R$ are obtained directly from optical readout. The standard factorization then follows as $A=QR$ with $Q=U^\dagger$ stored implicitly in the programmed mesh. Within the physical operation model adopted here, the overall cost scales as $O(N\log_2 N)$, contrasting with the $O(N^3)$ arithmetic complexity of conventional dense digital QR routines.

An important feature of the approach is that the same hardware supports not only a single QR decomposition, but also a broader family of structured linear algebra procedures. In particular, we described native optical implementations of QR iteration, Hessenberg reduction, and bidiagonalization. Since QR factorization is typically used as a component of eigensolvers and singular value algorithms, this suggests that programmable interferometric meshes can serve as physical processors for complete spectral workflows rather than only as accelerators for repeated matrix-vector multiplication.

Our comparison with systolic arrays clarifies both the advantages and the limitations of the photonic approach. For QR decomposition, a fully pipelined systolic architecture based on Givens rotations can achieve $O(N)$ latency, which is asymptotically better than the $O(N\log_2 N)$ scaling obtained here. This advantage relies on concurrent configuration and propagation within a digital pipeline. In blocked settings, both approaches have the same leading order complexity, $O(N^3/D^2)$, while the optical realization may still benefit from high carrier frequency and multiplexing across independent channels. For Hessenberg reduction and bidiagonalization, the situation is different: the optical protocols retain $O(N\log_2 N)$ complexity, whereas direct systolic realizations require $O(N^2)$ steps because sequential data transport through the array cannot be neglected.

The proposed construction also highlights an architectural tradeoff. A natural extension would be to implement the same ideas on a rectangular Clements mesh, which offers improved layout regularity and component density relative to the triangular Reck geometry \cite{clementsOptimalDesignUniversal2016}. At present, however, known self-configuration procedures for the rectangular mesh require $O(N^2)$ steps \cite{hamerlyAccurateSelfConfigurationRectangular2022}, so the configuration overhead would remove the main asymptotic advantage of the present protocol.

Although we have focused on integrated photonics, the underlying principles are not restricted to the optical domain. The same concentration based protocols can be transferred to radio frequency or microwave hardware using hybrid couplers and tunable phase shifters \cite{pozarMicrowaveEngineering2012}. Such platforms may offer complementary advantages, including higher power handling and mature control electronics, while preserving the same interferometric mechanism.

Several practical considerations define the regime in which the proposed approach is most compelling. The complexity estimates rely on fast local feedback, on treating vector injection and output readout as elementary physical operations, and on the efficient reuse of programmed interferometric settings. Within that regime, the method is particularly attractive when the matrix is available in physically encoded form and when the task at hand is dominated by decompositions or iterative spectral searches. Finally, while the presentation has been formulated for complex matrices, the same procedures apply directly to real valued matrices as a special case.

\begin{acknowledgments}
	S.A.F. acknowledges the support from the Foundation for the Advancement of Theoretical Physics and Mathematics (BASIS) (Project № 23-2-10-15-1) and the Scholarships of the President of the Russian Federation for postgraduate students.
	% {\color{red} S.A.F. work was conducted under the Russian Roadmap on Quantum Computing (Contract № 868-1.3-15/15-2021 dated 10/05/2021)}.
\end{acknowledgments}

\bibliography{biblio.bib}% Produces the bibliography via BibTeX.

@book{kung1978systolic,
	title = {Systolic Arrays for ({{VLSI}})},
	author = {Kung, H.T. and Leiserson, C.E. and {of COMPUTER SCIENCE.}, CARNEGIE-MELLON UNIV PITTSBURGH PA Dept. and Department, Carnegie Mellon University. Computer Science},
	year = 1978,
	series = {{{CMU-CS}}},
	publisher = {Carnegie-Mellon University, Department of Computer Science},
	url = {https://books.google.ru/books?id=pAKfHAAACAAJ}
}

@misc{lu2025systolicupdateschemeovercome,
	title={A systolic update scheme to overcome memory bandwidth limitations in GPU-accelerated FDTD simulations}, 
	author={Jesse Lu and David Qu and Jim Qu and Ryan Fong and Geun Ho Ahn and Jelena Vuckovic},
	year={2025},
	eprint={2502.20610},
	archivePrefix={arXiv},
	primaryClass={physics.optics}
}

@article{Feldmann_2021,
	title={Parallel convolutional processing using an integrated photonic tensor core},
	volume={589},
	ISSN={1476-4687},
	url={http://dx.doi.org/10.1038/s41586-020-03070-1},
	DOI={10.1038/s41586-020-03070-1},
	number={7840},
	journal={Nature},
	publisher={Springer Science and Business Media LLC},
	author={Feldmann, J. and Youngblood, N. and Karpov, M. and Gehring, H. and Li, X. and Stappers, M. and Le Gallo, M. and Fu, X. and Lukashchuk, A. and Raja, A. S. and Liu, J. and Wright, C. D. and Sebastian, A. and Kippenberg, T. J. and Pernice, W. H. P. and Bhaskaran, H.},
	year={2021},
	month=jan, pages={52–58} }

@misc{li2025allopticalcomputing100ghzclock,
	title={All-optical computing with beyond 100-GHz clock rates}, 
	author={Gordon H. Y. Li and Midya Parto and Jinhao Ge and Qing-Xin Ji and Maodong Gao and Yan Yu and James Williams and Robert M. Gray and Christian R. Leefmans and Nicolas Englebert and Kerry J. Vahala and Alireza Marandi},
	year={2025},
	eprint={2501.05756},
	archivePrefix={arXiv},
	primaryClass={physics.optics}, 
}

@article{kurzakDesignImplementationPULSAR2017,
	title = {Design and {{Implementation}} of the {{PULSAR Programming System}} for {{Large Scale Computing}}},
	author = {Kurzak, Jakub and Luszczek, Piotr and Yamazaki, Ichitaro and Robert, Yves and Dongarra, Jack},
	year = 2017,
	month = mar,
	journal = {Supercomputing Frontiers and Innovations},
	volume = {4},
	number = {1},
	issn = {23138734},
	doi = {10.14529/jsfi170101},
	url = {https://superfri.org/index.php/superfri/article/view/121},
	urldate = {2026-03-28}
}

@article{kungVLSIArrayProcessors1985,
	title = {{{VLSI Array}} Processors},
	author = {Kung, S.},
	year = 1985,
	journal = {IEEE ASSP Magazine},
	volume = {2},
	number = {3},
	pages = {4--22},
	issn = {0740-7467},
	doi = {10.1109/MASSP.1985.1163741},
	url = {http://ieeexplore.ieee.org/document/1163741/},
	urldate = {2026-03-20},
	copyright = {https://ieeexplore.ieee.org/Xplorehelp/downloads/license-information/IEEE.html}
}

@inproceedings{gentlemanMatrixTriangularizationSystolic1982,
	title = {Matrix {{Triangularization By Systolic Arrays}}},
	booktitle = {25th {{Annual Technical Symposium}}},
	author = {Gentleman, W. M. and Kung, H. T.},
	editor = {Tao, Tien F.},
	year = 1982,
	month = jul,
	pages = {19--26},
	address = {San Diego},
	doi = {10.1117/12.932507},
	url = {http://proceedings.spiedigitallibrary.org/proceeding.aspx?articleid=1231935},
	urldate = {2026-03-20}
}

@article{bojanczykNumericallyStableSolution1984,
	title = {Numerically {{Stable Solution}} of {{Dense Systems}} of {{Linear Equations Using Mesh-Connected Processors}}},
	author = {Bojanczyk, A. and Brent, R. P. and Kung, H. T.},
	year = 1984,
	month = mar,
	journal = {SIAM Journal on Scientific and Statistical Computing},
	volume = {5},
	number = {1},
	pages = {95--104},
	issn = {0196-5204, 2168-3417},
	doi = {10.1137/0905007},
	url = {http://epubs.siam.org/doi/10.1137/0905007},
	urldate = {2026-03-20}
}

@article{tiskinCommunicationefficientParallelGeneric2007,
	title = {Communication-Efficient Parallel Generic Pairwise Elimination},
	author = {Tiskin, Alexander},
	year = 2007,
	month = feb,
	journal = {Future Generation Computer Systems},
	volume = {23},
	number = {2},
	pages = {179--188},
	issn = {0167739X},
	doi = {10.1016/j.future.2006.04.017},
	url = {https://linkinghub.elsevier.com/retrieve/pii/S0167739X06000835},
	urldate = {2026-04-05},
	copyright = {https://www.elsevier.com/tdm/userlicense/1.0/}
}

@article{chilesDesignFabricationMetrology2018,
	title = {Design, Fabrication, and Metrology of 10 \texttimes{} 100 Multi-Planar Integrated Photonic Routing Manifolds for Neural Networks},
	author = {Chiles, Jeff and Buckley, Sonia M. and Nam, Sae Woo and Mirin, Richard P. and Shainline, Jeffrey M.},
	year = 2018,
	month = oct,
	journal = {APL Photonics},
	volume = {3},
	number = {10},
	pages = {106101},
	issn = {2378-0967},
	doi = {10.1063/1.5039641},
	urldate = {2025-12-28},
}

@article{giamougiannisNeuromorphicSiliconPhotonics2023,
	title = {Neuromorphic Silicon Photonics with 50 {{GHz}} Tiled Matrix Multiplication for Deep-Learning Applications},
	author = {Giamougiannis, George and Tsakyridis, Apostolos and {Moralis-Pegios}, Miltiadis and {Mourgias-Alexandris}, George and Totovic, Angelina R. and Dabos, George and Kirtas, Manos and Passalis, Nikolaos and Tefas, Anastasios and Kalavrouziotis, Dimitrios and Syrivelis, Dimitris and Bakopoulos, Paraskevas and Mentovich, Elad and Lazovsky, David and Pleros, Nikos},
	year = 2023,
	month = feb,
	journal = {Advanced Photonics},
	volume = {5},
	number = {1},
	pages = {016004},
	publisher = {SPIE},
	issn = {2577-5421, 2577-5421},
	doi = {10.1117/1.AP.5.1.016004},
	urldate = {2025-12-28},
}

@article{xuParallelOpticalCoherent2022,
	title = {Parallel Optical Coherent Dot-Product Architecture for Large-Scale Matrix Multiplication with Compatibility for Diverse Phase Shifters},
	author = {Xu, Shaofu and Wang, Jing and Yi, Sicheng and Zhao, Xinrui and Liu, Binshuo and Shao, Jiayi and Zou, Weiwen},
	year = 2022,
	month = nov,
	journal = {Optics Express},
	volume = {30},
	number = {23},
	pages = {42057--42068},
	publisher = {Optica Publishing Group},
	issn = {1094-4087},
	doi = {10.1364/OE.471519},
	urldate = {2025-12-28},
	copyright = {\copyright{} 2022 Optica Publishing Group},
}

@article{bogaertsProgrammablePhotonicCircuits2020,
	title = {Programmable Photonic Circuits},
	author = {Bogaerts, Wim and P{\'e}rez, Daniel and Capmany, Jos{\'e} and Miller, David A. B. and Poon, Joyce and Englund, Dirk and Morichetti, Francesco and Melloni, Andrea},
	year = 2020,
	month = oct,
	journal = {Nature},
	volume = {586},
	number = {7828},
	pages = {207--216},
	issn = {0028-0836, 1476-4687},
	doi = {10.1038/s41586-020-2764-0},
	url = {https://www.nature.com/articles/s41586-020-2764-0},
	urldate = {2026-04-05}
}

@article{wuProgrammableIntegratedPhotonic2024,
	title = {Programmable Integrated Photonic Coherent Matrix: {{Principle}}, Configuring, and Applications},
	shorttitle = {Programmable Integrated Photonic Coherent Matrix},
	author = {Wu, Bo and Zhou, Hailong and Dong, Jianji and Zhang, Xinliang},
	year = 2024,
	month = mar,
	journal = {Applied Physics Reviews},
	volume = {11},
	number = {1},
	pages = {011309},
	issn = {1931-9401},
	doi = {10.1063/5.0184982},
	urldate = {2025-12-28},
}

@article{kondratyevLargescaleErrortolerantProgrammable2024,
	title = {Large-Scale Error-Tolerant Programmable Interferometer Fabricated by Femtosecond Laser Writing},
	author = {Kondratyev, Ilya and Ivanova, Veronika and Fldzhyan, Suren and Argenchiev, Artem and Kostyuchenko, Nikita and Zhuravitskii, Sergey and Skryabin, Nikolay and Dyakonov, Ivan and Saygin, Mikhail and Straupe, Stanislav and Korneev, Alexander and Kulik, Sergei},
	year = 2024,
	month = mar,
	journal = {Photonics Research},
	volume = {12},
	number = {3},
	pages = {A28},
	issn = {2327-9125},
	doi = {10.1364/PRJ.504588},
	url = {https://opg.optica.org/abstract.cfm?URI=prj-12-3-A28},
	urldate = {2025-12-28}
}

@article{kondratyevEffectiveProgrammingPhotonic2026,
	title = {Effective Programming of a Photonic Processor with Complex Interferometric Structure},
	author = {Kondratyev, I.V. and Urusova, K.N. and Argenchiev, A.S. and Klushnikov, N.S. and Kuzmin, S.S. and Skryabin, N.N. and Golikov, A.D. and Kovalyuk, V.V. and Goltsman, G.N. and Dyakonov, I.V. and Straupe, S.S. and Kulik, S.P.},
	year = 2026,
	month = mar,
	journal = {Physical Review Applied},
	volume = {25},
	number = {3},
	pages = {034072},
	issn = {2331-7019},
	doi = {10.1103/btbd-w8fk},
	url = {https://link.aps.org/doi/10.1103/btbd-w8fk},
	urldate = {2026-03-31}
}

@article{jouppiInDatacenterPerformanceAnalysis2017,
	title = {In-{{Datacenter Performance Analysis}} of a {{Tensor Processing Unit}}},
	author = {Jouppi, Norman P. and Young, Cliff and Patil, Nishant and Patterson, David and Agrawal, Gaurav and Bajwa, Raminder and Bates, Sarah and Bhatia, Suresh and Boden, Nan and Borchers, Al and et. al.
	},
	year = 2017,
	month = sep,
	journal = {ACM SIGARCH Computer Architecture News},
	volume = {45},
	number = {2},
	pages = {1--12},
	issn = {0163-5964},
	doi = {10.1145/3140659.3080246},
	url = {https://dl.acm.org/doi/10.1145/3140659.3080246},
	urldate = {2026-03-31}
}

@article{pesericoIntegratedPhotonicTensor2023,
	title = {Integrated {{Photonic Tensor Processing Unit}} for a {{Matrix Multiply}}: {{A Review}}},
	shorttitle = {Integrated {{Photonic Tensor Processing Unit}} for a {{Matrix Multiply}}},
	author = {Peserico, Nicola and Shastri, Bhavin J. and Sorger, Volker J.},
	year = 2023,
	month = jun,
	journal = {Journal of Lightwave Technology},
	volume = {41},
	number = {12},
	pages = {3704--3716},
	issn = {0733-8724, 1558-2213},
	doi = {10.1109/JLT.2023.3269957},
	url = {https://ieeexplore.ieee.org/document/10114395/},
	urldate = {2026-01-23},
	copyright = {https://creativecommons.org/licenses/by/4.0/legalcode}
}

@misc{carmonaLuxIALightweightUnitary2025,
	title = {{{LuxIA}}: {{A Lightweight Unitary matriX-based Framework Built}} on an {{Iterative Algorithm}} for {{Photonic Neural Network Training}}},
	shorttitle = {{{LuxIA}}},
	author = {Carmona, Tzamn Melendez and Marchesin, Federico and Abrate, Marco P. and Bienstman, Peter and Carlo, Stefano Di and Senior, Alessandro Savino},
	year = 2025,
	month = dec,
	number = {arXiv:2512.22264},
	eprint = {2512.22264},
	primaryclass = {cs},
	publisher = {arXiv},
	archiveprefix = {arXiv}
}

@article{linRobustMZIBasedOptical2025,
	title = {A {{Robust MZI-Based Optical Neural Network Using QR Decomposition}}},
	author = {Lin, Jian and Yang, Kang and Fu, Qiang and Wang, Pengjun and Dai, Shixun and Chen, Weiwei and Kong, Dejun and Li, Jun and Dai, Tingge and Yang, Jianyi},
	year = 2025,
	month = feb,
	journal = {Journal of Lightwave Technology},
	volume = {43},
	number = {3},
	pages = {1024--1031},
	issn = {0733-8724, 1558-2213},
	doi = {10.1109/JLT.2024.3476113},
	url = {https://ieeexplore.ieee.org/document/10707275/},
	urldate = {2026-01-15},
	copyright = {https://ieeexplore.ieee.org/Xplorehelp/downloads/license-information/IEEE.html}
}

@misc{nemkovComplexityenergyTradeoffProgrammable2025,
	title = {Complexity-Energy Trade-off in Programmable Unitary Interferometers},
	author = {Nemkov, Nikita A. and Straupe, Stanislav S.},
	year = 2025,
	month = jul,
	number = {arXiv:2507.22972},
	eprint = {2507.22972},
	primaryclass = {physics},
	publisher = {arXiv},
	urldate = {2025-12-28},
	archiveprefix = {arXiv}
}

@article{taguchiSubquadraticScalableApproximate2025,
	title = {Sub-Quadratic Scalable Approximate Linear Converter Using Multi-Plane Light Conversion with Low-Entropy Mode Mixers},
	author = {Taguchi, Yoshitaka},
	year = 2025,
	month = oct,
	journal = {Journal of the Optical Society of America B},
	volume = {42},
	number = {10},
	pages = {2207},
	issn = {0740-3224, 1520-8540},
	doi = {10.1364/JOSAB.568104},
	url = {https://opg.optica.org/abstract.cfm?URI=josab-42-10-2207},
	urldate = {2025-12-28}
}

@article{hamerlyAsymptoticallyFaulttolerantProgrammable2022,
	title = {Asymptotically Fault-Tolerant Programmable Photonics},
	author = {Hamerly, Ryan and Bandyopadhyay, Saumil and Englund, Dirk},
	year = 2022,
	month = nov,
	journal = {Nature Communications},
	volume = {13},
	number = {1},
	pages = {6831},
	issn = {2041-1723},
	doi = {10.1038/s41467-022-34308-3},
	url = {https://www.nature.com/articles/s41467-022-34308-3},
	urldate = {2025-12-28}
}

@article{hamerlyAccurateSelfConfigurationRectangular2022,
	title = {Accurate {{Self-Configuration}} of {{Rectangular Multiport Interferometers}}},
	author = {Hamerly, Ryan and Bandyopadhyay, Saumil and Englund, Dirk},
	year = 2022,
	month = aug,
	journal = {Physical Review Applied},
	volume = {18},
	number = {2},
	pages = {024019},
	issn = {2331-7019},
	doi = {10.1103/PhysRevApplied.18.024019},
	url = {https://link.aps.org/doi/10.1103/PhysRevApplied.18.024019},
	urldate = {2025-12-28}
}

@article{hamerlyInformationtheoreticLimitProgrammable2025,
	title = {Toward the Information-Theoretic Limit of Programmable Photonics},
	author = {Hamerly, Ryan and Basani, Jasvith R. and Sludds, Alexander and Vadlamani, Sri K. and Englund, Dirk},
	year = 2025,
	month = nov,
	journal = {APL Photonics},
	volume = {10},
	number = {11},
	pages = {110803},
	issn = {2378-0967},
	doi = {10.1063/5.0269741},
	url = {https://pubs.aip.org/app/article/10/11/110803/3370635/Toward-the-information-theoretic-limit-of},
	urldate = {2025-12-28}
}

@article{talibPhotonicMatrixMultiplication2025,
	title = {Photonic Matrix Multiplication Circuit Based on Double Racetrack Resonator Building Blocks},
	author = {Talib, Hussein and Sewell, Phillip D. and Vukovic, Ana and Phang, Sendy},
	year = 2025,
	month = oct,
	journal = {Optical and Quantum Electronics},
	volume = {57},
	number = {11},
	pages = {590},
	issn = {1572-817X},
	doi = {10.1007/s11082-025-08524-2},
	url = {https://doi.org/10.1007/s11082-025-08524-2},
	urldate = {2026-01-08}
}

@article{girouardNearoptimalDecompositionUnitary2026,
	title = {Near-Optimal Decomposition of Unitary Matrices Using Phase Masks and the Discrete {{Fourier}} Transform},
	author = {Girouard, Vincent and Quesada, Nicol{\'a}s},
	year = 2026,
	month = mar,
	journal = {Journal of the Optical Society of America B},
	volume = {43},
	number = {3},
	pages = {A66},
	issn = {0740-3224, 1520-8540},
	doi = {10.1364/JOSAB.577579}
}

@article{chenIterativePhotonicProcessor2022,
	title = {Iterative Photonic Processor for Fast Complex-Valued Matrix Inversion},
	author = {Chen, Minjia and Cheng, Qixiang and Ayata, Masafumi and Holm, Mark and Penty, Richard},
	year = 2022,
	month = nov,
	journal = {Photonics Research},
	volume = {10},
	number = {11},
	pages = {2488},
	issn = {2327-9125},
	doi = {10.1364/PRJ.468097},
	url = {https://opg.optica.org/abstract.cfm?URI=prj-10-11-2488},
	urldate = {2025-12-28}
}

@article{chenOefficientIterativeMatrix2024,
	title = {I/{{O-efficient}} Iterative Matrix Inversion with Photonic Integrated Circuits},
	author = {Chen, Minjia and Wang, Yizhi and Yao, Chunhui and Wonfor, Adrian and Yang, Shuai and Penty, Richard and Cheng, Qixiang},
	year = 2024,
	month = jul,
	journal = {Nature Communications},
	volume = {15},
	number = {1},
	pages = {5926},
	publisher = {Nature Publishing Group},
	issn = {2041-1723},
	doi = {10.1038/s41467-024-50302-3},
	url = {https://www.nature.com/articles/s41467-024-50302-3},
	urldate = {2026-01-08},
	copyright = {2024 The Author(s)}
}

@inproceedings{milanizadehRecursiveMZIMesh2020,
	title = {Recursive {{MZI}} Mesh for Integral Equation Implementation},
	booktitle = {European {{Conference}} on {{Integrated Optics}} 2020 ({{ECIO}})},
	author = {Milanizadeh, Maziyar and Damiani, Elena and Jonuzi, Tigers and Mencagli, Mario Junior and Edwards, Brian and Miller, David AB and Engheta, Nader and Melloni, Andrea and Morichetti, Francesco},
	year = 2020,
	url = {https://www-ee.stanford.edu/\textasciitilde dabm/466.pdf},
	urldate = {2026-01-08}
}

@inproceedings{cavicchioliProgrammableIntegratedPhotonic2024,
	title = {Programmable Integrated Photonic Circuit for Matrix Inversion},
	booktitle = {Optical {{Fiber Communication Conference}}},
	author = {Cavicchioli, G. and Miller, D. A. B. and Engheta, N. and Melloni, A. and Morichetti, F.},
	year = 2024,
	pages = {Th1A--2},
	publisher = {Optica Publishing Group},
	url = {https://opg.optica.org/abstract.cfm?uri=ofc-2024-Th1A.2},
	urldate = {2026-01-08}
}

@article{xiaoResidualCalibrationHighprecision2025,
	title = {Residual Calibration for High-Precision Optical Neural Networks},
	author = {Xiao, Yao and Zhao, Yang and Wang, Wei and Cheng, Zhitao and Peng, Xizhu and Tang, He and Liu, Shengping and Tang, Yong},
	year = 2025,
	month = jul,
	journal = {Optics Express},
	volume = {33},
	number = {15},
	pages = {32190},
	issn = {1094-4087},
	doi = {10.1364/OE.564828},
	url = {https://opg.optica.org/abstract.cfm?URI=oe-33-15-32190},
	urldate = {2026-01-08}
}

@misc{brugiereNewDesignsLinear2025,
	title = {New Designs of Linear Optical Interferometers with Minimal Depth and Component Count},
	author = {de Brugi{\`e}re, Timoth{\'e}e Goubault and Mezher, Rawad and Currie, Sebastian and Mansfield, Shane},
	year = 2025,
	month = apr,
	number = {arXiv:2504.06059},
	eprint = {2504.06059},
	primaryclass = {quant-ph},
	publisher = {arXiv},
	urldate = {2025-12-28},
	archiveprefix = {arXiv}
}

@book{tyrtyshnikovBriefIntroductionNumerical1997,
	title = {A {{Brief Introduction}} to {{Numerical Analysis}}},
	author = {Tyrtyshnikov, Eugene E.},
	year = 1997,
	publisher = {Birkh\"auser Boston},
	address = {Boston, MA},
	doi = {10.1007/978-0-8176-8136-4},
	urldate = {2025-12-28},
	copyright = {http://www.springer.com/tdm},
	isbn = {978-1-4612-6413-2 978-0-8176-8136-4},
}

@book{arbenzLectureNotesSolving2016,
	title = {Lecture {{Notes}} on {{Solving Large Scale Eigenvalue Problems}}},
	author = {Arbenz, Dr Peter},
	year = 2016,
	publisher = {Computer Science Department, ETH Z\"urich},
	url = {https://people.inf.ethz.ch/arbenz/ewp/lnotes.html}
}

@article{marchesinBraidedInterferometerMesh2025,
	title = {Braided Interferometer Mesh for Robust Photonic Matrix-Vector Multiplications with Non-Ideal Components},
	author = {Marchesin, Federico and Hejda, Mat{\v e}j and Carmona, Tzamn Melendez and Carlo, Stefano Di and Savino, Alessandro and Pavanello, Fabio and Vaerenbergh, Thomas Van and Bienstman, Peter},
	year = 2025,
	month = jan,
	journal = {Optics Express},
	volume = {33},
	number = {2},
	primaryclass = {physics},
	pages = {2227},
	issn = {1094-4087},
	doi = {10.1364/OE.547206}
}

@book{golubMatrixComputations2013,
	title = {Matrix Computations},
	author = {Golub, Gene H. and Van Loan, Charles F.},
	year = 2013,
	series = {Johns {{Hopkins}} Studies in the Mathematical Sciences},
	edition = {4},
	publisher = {The Johns Hopkins University Press},
	address = {Baltimore},
	url = {https://doi.org/10.56021/9781421407944},
	isbn = {978-1-4214-0794-4}
}

@article{ditriaHighPrecisionAutomatedSetting2025,
	title = {High-{{Precision Automated Setting}} of {{Arbitrary Magnitude}} and {{Phase}} of {{Mach}}--{{Zehnder Interferometers}} for {{Scalable Optical Computing}}},
	author = {Di Tria, Alessandro and Cavicchioli, Gabriele and Giannoccaro, Pietro and Morichetti, Francesco and Melloni, Andrea and Ferrari, Giorgio and Sampietro, Marco and Zanetto, Francesco},
	year = 2025,
	month = sep,
	journal = {Laser \& Photonics Reviews},
	pages = {e00610},
	issn = {1863-8880, 1863-8899},
	doi = {10.1002/lpor.202500610},
	url = {https://onlinelibrary.wiley.com/doi/10.1002/lpor.202500610},
	urldate = {2026-01-09}
}

@article{tangTwolayerIntegratedPhotonic2022,
	title = {Two-Layer Integrated Photonic Architectures with Multiport Photodetectors for High-Fidelity and Energy-Efficient Matrix Multiplications},
	author = {Tang, Rui and Okano, Makoto and Toprasertpong, Kasidit and Takagi, Shinichi and Englund, Dirk and Takenaka, Mitsuru},
	year = 2022,
	month = sep,
	journal = {Optics Express},
	volume = {30},
	number = {19},
	pages = {33940},
	issn = {1094-4087},
	doi = {10.1364/OE.457258},
	url = {https://opg.optica.org/abstract.cfm?URI=oe-30-19-33940},
	urldate = {2025-12-28}
}

@article{tangLowerdepthProgrammableLinear2024,
	title = {Lower-Depth Programmable Linear Optical Processors},
	author = {Tang, Rui and Tanomura, Ryota and Tanemura, Takuo and Nakano, Yoshiaki},
	year = 2024,
	month = jan,
	journal = {Physical Review Applied},
	volume = {21},
	number = {1},
	pages = {014054},
	issn = {2331-7019},
	doi = {10.1103/PhysRevApplied.21.014054},
	url = {https://link.aps.org/doi/10.1103/PhysRevApplied.21.014054},
	urldate = {2025-12-28}
}

@article{millerSelfconfiguringUniversalLinear2013,
	title = {Self-Configuring Universal Linear Optical Component},
	author = {Miller, David A B},
	year = 2013,
	month = jun,
	journal = {Photonics Research},
	volume = {1},
	number = {1},
	pages = {1},
	doi = {10.1364/PRJ.1.000001}
}

@article{tangWaveguidemultiplexedPhotonicMatrix2025,
	title = {Waveguide-Multiplexed Photonic Matrix--Vector Multiplication Processor Using Multiport Photodetectors},
	author = {Tang, Rui and Okano, Makoto and Zhang, Chao and Toprasertpong, Kasidit and Takagi, Shinichi and Takenaka, Mitsuru},
	year = 2025,
	month = jun,
	journal = {Optica},
	volume = {12},
	number = {6},
	pages = {812},
	issn = {2334-2536},
	doi = {10.1364/OPTICA.552023},
	url = {https://opg.optica.org/abstract.cfm?URI=optica-12-6-812},
	urldate = {2025-12-28}
}

@article{kuzminLeveragingFeaturebasedModel2025,
	title = {Leveraging a Feature-Based Model for Linear Optical Interferometer Control},
	author = {Kuzmin, Sergei and Dyakonov, Ivan and Straupe, Stanislav},
	year = 2025,
	month = nov,
	journal = {Physical Review A},
	volume = {112},
	number = {5},
	pages = {053515},
	issn = {2469-9926, 2469-9934},
	doi = {10.1103/x8wd-83pt},
	url = {https://link.aps.org/doi/10.1103/x8wd-83pt},
	urldate = {2025-12-28}
}

@article{fldzhyanLowdepthTwounitaryDesign2026,
	title = {Low-Depth Two-Unitary Design of Programmable Photonic Circuits},
	author = {Fldzhyan, S. A. and Saygin, M. {\relax Yu}. and Straupe, S. S.},
	year = 2026,
	month = jan,
	journal = {Physical Review Research},
	volume = {8},
	number = {1},
	pages = {013021},
	issn = {2643-1564},
	doi = {10.1103/9fg1-fdfl},
	url = {https://link.aps.org/doi/10.1103/9fg1-fdfl},
	urldate = {2026-01-12}
}

@article{bantyshFastReconstructionProgrammable2023,
	title = {Fast Reconstruction of Programmable Integrated Interferometers},
	author = {Bantysh, Boris and Katamadze, Konstantin and Chernyavskiy, Andrey and Bogdanov, Yurii},
	year = 2023,
	month = may,
	journal = {Optics Express},
	volume = {31},
	number = {10},
	pages = {16729},
	issn = {1094-4087},
	doi = {10.1364/OE.487156},
	url = {https://opg.optica.org/abstract.cfm?URI=oe-31-10-16729},
	urldate = {2025-12-28}
}

@article{bantyshFastReconstructionProgrammable2024,
	title = {Fast Reconstruction of Programmable Interferometers with Intensity-Only Measurements},
	author = {Bantysh, B I and Chernyavskiy, A Yu and Fldzhyan, S A and Bogdanov, Yu I},
	year = 2024,
	month = jan,
	journal = {Laser Physics Letters},
	volume = {21},
	number = {1},
	pages = {015203},
	issn = {1612-2011, 1612-202X},
	doi = {10.1088/1612-202X/ad0caf},
	url = {https://iopscience.iop.org/article/10.1088/1612-202X/ad0caf},
	urldate = {2025-12-28}
}

@ARTICLE{2020SciPy-NMeth,
	author  = {Virtanen, Pauli 
	and Gommers, Ralf and Oliphant, Travis E. and
	Haberland, Matt and Reddy, Tyler and Cournapeau, David and
	Burovski, Evgeni and Peterson, Pearu and Weckesser, Warren and
	Bright, Jonathan and et. al.},
	title   = {{{SciPy} 1.0: Fundamental Algorithms for Scientific
	Computing in Python}},
	journal = {Nature Methods},
	year    = {2020},
	volume  = {17},
	pages   = {261--272},
	adsurl  = {https://rdcu.be/b08Wh},
	doi     = {10.1038/s41592-019-0686-2},
}

@article{khachaturianIQPhotonicReceiver2021,
	title = {{{IQ Photonic Receiver}} for {{Coherent Imaging With}} a {{Scalable Aperture}}},
	author = {Khachaturian, Aroutin and Fatemi, Reza and Hajimiri, Ali},
	year = 2021,
	journal = {IEEE Open Journal of the Solid-State Circuits Society},
	volume = {1},
	pages = {263--270},
	issn = {2644-1349},
	doi = {10.1109/OJSSCS.2021.3113264},
	url = {https://ieeexplore.ieee.org/document/9540747},
	urldate = {2026-01-07}
}

@article{russellDirectDiallingHaar2017,
	title = {Direct Dialling of {{Haar}} Random Unitary Matrices},
	author = {Russell, Nicholas J and Chakhmakhchyan, Levon and O'Brien, Jeremy L and Laing, Anthony},
	year = 2017,
	month = mar,
	journal = {New Journal of Physics},
	volume = {19},
	number = {3},
	pages = {033007},
	issn = {1367-2630},
	doi = {10.1088/1367-2630/aa60ed},
	urldate = {2025-12-28},
}

@article{reckExperimentalRealizationAny1994,
	title = {Experimental Realization of Any Discrete Unitary Operator},
	author = {Reck, Michael and Zeilinger, Anton and Bernstein, Herbert J. and Bertani, Philip},
	year = 1994,
	month = jul,
	journal = {Physical Review Letters},
	volume = {73},
	number = {1},
	pages = {58--61},
	issn = {0031-9007},
	doi = {10.1103/PhysRevLett.73.58},
	urldate = {2025-12-28},
	copyright = {http://link.aps.org/licenses/aps-default-license},
}

@article{pottonReciprocityOptics2004,
	title = {Reciprocity in Optics},
	author = {Potton, R. J.},
	year = 2004,
	month = apr,
	journal = {Reports on Progress in Physics},
	volume = {67},
	number = {5},
	pages = {717},
	issn = {0034-4885},
	doi = {10.1088/0034-4885/67/5/R03},
	url = {https://doi.org/10.1088/0034-4885/67/5/R03},
	urldate = {2026-01-08}
}

@article{clementsOptimalDesignUniversal2016,
	title = {Optimal Design for Universal Multiport Interferometers},
	author = {Clements, William R. and Humphreys, Peter C. and Metcalf, Benjamin J. and Kolthammer, W. Steven and Walsmley, Ian A.},
	year = 2016,
	month = dec,
	journal = {Optica},
	volume = {3},
	number = {12},
	pages = {1460},
	issn = {2334-2536},
	doi = {10.1364/OPTICA.3.001460},
	url = {https://opg.optica.org/abstract.cfm?URI=optica-3-12-1460},
	urldate = {2025-12-28},
	copyright = {https://creativecommons.org/licenses/by/4.0/}
}

@book{pozarMicrowaveEngineering2012,
	title = {Microwave Engineering},
	author = {Pozar, David M.},
	year = 2012,
	edition = {Fourth edition},
	publisher = {John Wiley \& Sons, Inc},
	address = {Hoboken, NJ},
	url = {https://hajaress.wordpress.com/wp-content/uploads/2019/09/microwave_engineering_david_m_pozar_4ed_wiley_2012.pdf},
	isbn = {978-0-470-63155-3 978-1-118-21363-6}
}
% \printbibliography

% \onecolumngrid

\appendix

% \section{Modular Scaling}\label{sec:Appendix}
% \sout{
	% Current photonic chips support only a few dozen channels. To scale, we propose tiling: interconnecting smaller $d \times d$ tiles to form larger systems (Fig.~{\ref{fig:A}}).}

% \sout{For the unitary mesh (Fig.~{\ref{fig:A}}(a)), full parallel tiling with $\sim N^2/(2d^2)$ tiles preserves the $O(N^2)$ complexity of Section~{\ref{sec:QR dec}}. Alternatively, with a single tile reused sequentially, each placement configures $O(d^2)$ elements, while injecting and reading $N\cdot d$ amplitudes for each column. Since the full mesh has $\sim N^2/2$ $T$ blocks, total cost is}
% \begin{equation}
	% O\left(\frac{N^2}{d^2}(Nd + d^2)\right) = O\left(\frac{N^3}{d} + N^2\right).
	% \end{equation}
% \sout{Choosing $d \sim N^\alpha$ ($0 < \alpha \leq 1$) yields $O(N^{3-\alpha})$ --- better than the $O(N^3)$ digital baseline.}

% \sout{The X-bar stage (Fig.~{\ref{fig:A}}(b)) also tiles naturally {\cite{nemkovComplexityenergyTradeoffProgrammable2025}}:  $N$ columns encoding and fan-ins each scale as $O(N)$, and structured inputs (e.g., tridiagonal) can reduce this further.}

% \begin{figure}[H]
	%     \centering
	%     \vspace{10pt}
	%     \includegraphics[width=\linewidth]{Figs/Conquer.pdf}
	%     \caption{\sout{Tiling for scalability. (a) Unitary mesh tiling ($N=18$, $d=6$). (b) X-bar tiling with column encoding and fan-in. Skew indicates timing alignment needs at $T$ blocks.}}
	%     \label{fig:A}
	% \end{figure}
	\end{document}